\documentstyle[12pt]{article}
\begin{document}

\setcounter{page}{0}
\thispagestyle{empty}

\vspace*{-1in}
\begin{flushright}
CUPP-95/3\\
\end{flushright}
\vskip 50pt
\begin{center}
{\large\bf{ACCELERATOR, REACTOR AND ATMOSPHERIC}\\
\large\bf {NEUTRINO DATA: A THREE FLAVOUR OSCILLATION ANALYSIS }}\\

\vskip 25pt
{ \it Srubabati Goswami $^{a}$, Kamales Kar $^{b}$, Amitava
Raychaudhuri $^{a}$ \\
$^{a}$ Department of Pure Physics,\\University of Calcutta,\\
92 Acharya Prafulla Chandra Road,\\Calcutta 700 009, INDIA.\\
$^{b}$ Saha Institute of Nuclear Physics,\\1/AF, Bidhannagar
\\Calcutta 700 064, INDIA.\\}

\vskip 15pt

P.A.C.S. Nos.: 14.60.Pq, 14.60.Lm, 96.40.Tv

{\bf ABSTRACT}
\end{center}

\vskip 15pt

We perform a three flavour analysis of the atmospheric, accelerator
and reactor neutrino data from the Kamiokande, LSND and Bugey
experiments respectively. Choosing the values of ${\Delta{m}}^2$
obtained from two flavour fits of the first two experiments, the
allowed ranges of the three generation mixing angles are
determined.  The accelerator experiments CHORUS and NOMAD are found
to be most sensitive to regions of the allowed parameter space
which correspond to genuine three generation solutions for the
atmospheric neutrino anomaly.

\vskip 15pt
\parindent 0pt
May 24, 1995

\parindent 30pt

\newpage
In the standard model of electroweak theory the neutrinos are
assumed to be massless. But there is no compelling theoretical
reason behind this assumption. If the neutrinos are massive then,
as in the quark sector, the weak interaction basis of neutrinos may
be different from the mass eigenstate basis -- leading to mixing
between different flavours. A way for probing such mixing and small
neutrino masses is provided by neutrino oscillations. Two well
known neutrino puzzles that can be explained by flavour oscillation
of neutrinos are the solar neutrino problem and the atmospheric
neutrino anomaly. The recent declaration by the Liquid Scintillator
Neutrino Detector (LSND) collaboration \cite{lsnd} that they are
observing an excess of $\overline{\nu}_e$s (over the expected
background) which can be attributed to $\overline{\nu}_{\mu} -
\overline{\nu}_e$ oscillations has added a new dimension to the
issue of neutrino mass and mixing. LSND is most sensitive to
${\Delta m}^{2} \sim 6 {eV}^{2}$
\cite{caldwell} and the significance of this result for particle
physics, astrophysics and cosmology has been investigated
\cite{primack,silk}.  One notes that the three phenomena mentioned
above --  namely, the solar neutrino problem, the atmospheric
neutrino anomaly and the $\overline{\nu}_{\mu} - \overline{\nu}_e$
oscillations observed by the LSND group --  require vastly
different mass ranges.  The solar neutrino problem can be explained
either by Mikheyev-Smirnov-Wolfenstein oscillation \cite{msw} for
${{\Delta}m}^2 \sim 6 \times 10^{-6} eV^2$ and $\sin^{2}2\theta
\sim 7
\times 10^{-3}$ (non-adiabatic solution) and ${{\Delta}m}^2 \sim 9
\times 10^{-6} eV^2$ and $\sin^{2}2\theta \sim 0.6$ (large mixing angle
solution) \cite{langacker} or by oscillation in vacuum for
${{\Delta}m}^2 \sim (0.45-1.2) \times
10^{-10} eV^2$ and $\sin^{2}2\theta \sim (0.6-1.0)$ \cite{hata}
in a two generation scenario.
The atmospheric anomaly can be explained by either
$\nu_{\mu} - \nu_e$ or $\nu_{\mu} - \nu_{\tau}$ oscillations in a
two generation picture. The analysis of the new multi-GeV data
as well as the previous sub-GeV data of the
Kamiokande collaboration predicts the following best-fit
parameters (${\Delta m}^{2}, \sin^2 2\theta$) = (1.8 $\times 10^{-2}
eV^{2}, 1.0)$  for $\nu_{\mu} - \nu_e$ oscillation and
(1.6 $\times 10^{-2}eV^{2}, 1.0)$ for $\nu_{\mu} - \nu_{\tau}$
oscillation \cite{fukuda}.

Although each experiment can be explained by two flavour neutrino
oscillations, there are several motivations to go beyond this
approximation. The LEP result that there are three light neutrinos
is also supported by the requirements of nucleosynthesis in the
early universe. In the quark sector, mixing between three
generations is well established. A natural question then is how do
experiments constrain three neutrino mixing?  We stress that in a
realistic three flavour framework it is important to do a combined
analysis to find out the allowed range of parameters rather than
using separate two flavour schemes.  In particular, this might
uncover regions in the parameter space sensitive to the presence of
the third generation which cannot be probed in the two flavour
limit.

In this paper we perform a three flavour analysis of the
atmospheric and LSND data assuming that the presently reported
values will not change significantly as more results accumulate.
The constraints obtained from the reactor experiments are also
incorporated.  We take the  $\Delta{m}^2$s as: $\Delta_{12} \simeq
\Delta_{13} = 6 eV^{2}$ in the LSND range and $\Delta_{23} = 10^{-2}
{eV}^{2}$ as preferred by the atmospheric neutrino data. The
$\simeq$ sign means we neglect 10$^{-2}$ as compared to 6.  It will
become clear as we proceed that most of our analysis does not
depend on this specific choice as long as the order of magnitude
remains the same.
The three mixing angles are allowed to cover the whole range from 0
to $\pi$/2.  For atmospheric neutrinos, we find, in addition to the
two flavour results, genuine three generation solutions where both
$\nu_{\mu} - \nu_{e}$ and $\nu_{\mu} - \nu_{\tau}$ oscillation
channels simultaneously contribute.  The implications of the
allowed areas thus obtained for the accelerator experiments CHORUS
and NOMAD searching for $\nu_{\mu} -
\nu_{\tau}$ osillations are also discussed.
Such an  analysis for constraining the mixing angles has been
performed in
\cite{minakata} under the approximation of an effective two flavour
interpretation of the atmospheric neutrino problem either in the
$\nu_{\mu} - \nu_{e}$ or $\nu_{\mu} - \nu_{\tau}$ oscillation mode,
instead of a full three flavour investigation. A
detailed analysis combining the accelerator, reactor, solar and
atmospheric neutrino data had been carried out earlier (pre-LSND)
\cite{fogli1} taking a different spectrum for $\Delta{m}^2$ and
assuming the mixing angles to be less than $\pi$/4.

The measurement of atmospheric neutrino fluxes is being carried
out by the following groups --  Kamiokande, IMB, Fr\'{e}jus, Nusex and
Soudan2. So far, data of most statistical significance
have been collected by the Kamiokande and the IMB
collaborations. To reduce the uncertainty in the absolute flux
values the usual practice is to present the
ratio of ratios R which is defined as,
\begin{equation}
R = {\frac{(\nu_\mu + \overline{\nu}_{\mu})/
(\nu_e + \overline{\nu}_e)_{\rm obsvd}}
{(\nu_\mu + \overline{\nu}_{\mu})/(\nu_e + \overline{\nu}_e)_{\rm MC}}}
\label{ratm}
\end{equation}
where {MC} denotes the Monte-Carlo simulated ratio.
For neutrinos of energy less than $\sim$ 1 GeV, IMB finds R = $0.54
\pm 0.05 \pm 0.12$ \cite{imb} in agreement with the Kamiokande data
R = $0.60^{+0.06}_{-0.05} \pm 0.05$ in this energy range
\cite{fukuda,hirata}. Recently the Kamiokande collaboration has
published the results of the mesurement of the flux ratio in the
multi-GeV energy range \cite{fukuda}. They found R =
$0.57^{+0.08}_{-0.07} \pm 0.07$ in good agreement with the sub-GeV
value.  All these data show that R is smaller than the expected
value of unity, a result that might be explained by neutrino
flavour oscillation \cite{fogli2}. Another aspect of this
measurement that can independently point towards neutrino
oscillation is the dependence of R on the zenith-angle. The
multi-GeV Kamiokande data reveals a dependence on the zenith-angle
unlike the sub-GeV data, though the statistical significance of
this result has been questioned \cite{cern}. For the purpose of
this paper we use the sub-GeV Kamiokande results.

The LSND group searches for  $\overline{\nu}_{\mu} \rightarrow
\overline{\nu}_e$ oscillations using $\overline{\nu}_e$ appearance.
The $\overline{\nu}_e$s produce neutrons {\it via} the reaction
$\overline{\nu}_e p \rightarrow e^{+} n $ which in turn are
captured by protons producing a 2.2 MeV $\gamma$.  An excess of
beam-on events with a $\gamma$ of the above energy in time and
space coincidence with an electron in the energy range 36 MeV $ <
E_{e} < $ 60 MeV is considered as a signal for $\overline{\nu}_e$.
The mean source-detector distance is 30 metres.  The initial LSND
data reports an excess of ${16.4}^{+9.7}_{-8.9} \pm$ 3.3 events
over the estimated background which, if interpreted in terms of
neutrino oscillations, corresponds to a probability
$P_{\overline{\nu}_{\mu}\overline{\nu}_e}$ of
$({0.34}^{+0.20}_{-0.18} \pm 0.07)$\%.

Other appearance experiments searching for $\overline{\nu}_{\mu}
\rightarrow \overline{\nu}_e$ oscillations are KARMEN at the ISIS
spallation neutron facility \cite{karmen} and the  BNL-E776
\cite{bnl}. These experiments are consistent with no neutrino
oscillation.  KARMEN has so far quoted an upper limit on the
oscillation probability as
$P_{\overline{\nu}_{\mu}\overline{\nu}_e} \leq 3.1
\times 10^{-3}$ (90\% C.L.) wheareas from the two flavour exclusion
areas presented by BNL one gets
$P_{\overline{\nu}_{\mu}\overline{\nu}_e} \leq 1.5
\times 10^{-3}$ (90\% C.L.). In ref. \cite{lsnd}
the LSND group has shown that some of the areas allowed by
them in a two flavour analysis are disfavoured by KARMEN and
BNL-E776. In this paper we confine ourselves to the LSND data for
constraining the paremeters.

Reactor experiments searching for neutrino oscillation are
G\"{O}SGEN, KRASNOYARSK and Bugey. These experiments provide
exclusion regions in the ${\Delta{m}}^2$ - $\sin{^2}2\theta$
parameter space by non-observance of neutrino oscillation. The
maximum exclusion is by Bugey which measures the spectrum of
$\overline{\nu}_e$, coming from the Pressurized Water Reactors
running at the Bugey nuclear power plant, at 15, 40, and 95 metres
using neutron detection techniques. The 90\% C.L. exclusion contour
implies $1 - P_{\overline{\nu}_e\overline{\nu}_e} \leq $ 0.05
\cite{bugey}.

The general expression for the probability that an initial
$\nu_{\alpha}$ of energy $E$ gets converted to a $\nu_{\beta}$
after travelling a distance $L$ in vacuum is
\begin{equation}
P_{\nu_{\alpha}\nu_{\beta}} = \delta_{\alpha \beta} - 4~ \Sigma_{j
> i}~ U_{\alpha i} U_{\beta i} U_{\alpha j} U_{\beta j}
\sin^{2}(\frac{\pi L}{\lambda_{ij}})
\label{pab}
\end{equation}
where $\lambda_{ij} = 2.47m ({E_{\nu}}/{MeV})
({{eV}^{2}}/{\Delta_{ij}}$), $\Delta_{ij} = {m_j}^2 - {m_i}^2$.
The actual forms of the various survival and transition
probabilities depend on the spectrum of ${\Delta m}^{2}$ assumed
and the choice of the mixing matrix $U$ relating the flavour
eigenstates to the mass eigenstates.  The most suitable
parametrisation of $U$ for the mass spectrum chosen by us is $U =
R_{13} R_{12} R_{23}$ where $R_{ij}$ denotes the rotation matrix in
the $ij$-plane. This yields:
\begin{equation}
U = {\pmatrix {c_{12}c_{13} & s_{12}c_{13}c_{23} - s_{13}s_{23}
& c_{13}s_{12}s_{23} + s_{13}c_{23} \cr
-s_{12} & c_{12}c_{23} &c_{12}s_{23} \cr
-s_{13}c_{12} & -s_{13}s_{12}c_{23} - c_{13}s_{23} &
-s_{12}s_{13}s_{23} + c_{13}c_{23}\cr}}
\label{um}
\end{equation}
where $c_{ij} =\cos{{\theta}_{ij}}$ and $s_{ij} =\sin{{\theta}_{ij}}$.
We have assumed CP-invariance so that $U$ is real.
The above choice of $U$ has the advantage that ${\theta}_{23}$
does not appear in the expressions for the probability for LSND and
Bugey.
We now turn to the implications of the above mixing matrix and the
chosen mass ranges on the various probabilities.\\

\noindent{(i) LSND}\\
In order to see the impact of three neutrino generations, we
first note that for the energy and length scales relevant for LSND
$\lambda_{23} >> L$ and the term involving
$\sin^{2}(\pi L / \lambda_{23}) \rightarrow$ 0. Further,
$\lambda_{13} \simeq \lambda_{12}$ and
(\ref{pab}) simplifies to
\begin{equation}
P_{\overline{\nu}_{\mu}\overline{\nu}_e} = 4 c_{12}^2 s_{12}^2 c_{13}^2
{\sin^2}({\pi L/\lambda_{12}})
\label{pnumul}
\end{equation}

\noindent{(ii) Bugey}\\
For Bugey, the neutrino energy ranges from 2.8 - 7.8 MeV
whereas $L$ is typically $\sim$ 40 metres. Then $\lambda_{23} >> L$,
so that $\sin^{2}({\pi L}/{\lambda_{23}}) \rightarrow$ 0. On the
other hand $\lambda_{12} = \lambda_{13} << L$ so that
${\sin^2}(\pi L/\lambda_{12})$ and ${\sin^2}(\pi L/\lambda_{13})$
average out to 1/2. Then the relevant probability
is
\begin{equation}
P_{\overline{\nu}_e\overline{\nu}_e} = 1 - 2c_{13}^2c_{12}^2 +
2c_{13}^4c_{12}^4
\label{bugey}
\end{equation}

\noindent{(iii) Atmospheric neutrinos}\\
In a three flavour mixing scheme (\ref{ratm}) is given in terms of
the neutrino transition and survival probabilities as
\begin{equation}
R = { \frac{ P_{\nu_\mu \nu_\mu} + r_{MC} P_{\nu_\mu
\nu_e}}{P_{\nu_e \nu_e} + \frac{1}{r_{MC}} P_{\nu_\mu \nu_e}}}
\label{r3}
\end{equation}
where $r_{MC}$ = $(\nu_e + \overline{\nu}_e)/(\nu_{\mu} +
\overline{\nu}_{\mu})$ as obtained from a Monte-Carlo simulation.
Notice
that  for neutrinos in the energy range $\sim$ (0.1 -- 1) GeV
travelling through a distance ranging from $\sim$ (10 -- $10^4$) km,
$\lambda_{12} = \lambda_{13} << L$ and
${\sin^2}(\pi L/\lambda_{12})$ and ${\sin^2}(\pi L/\lambda_{13})$
can be replaced by their average value 1/2. Taking this into
account, the probabilities appearing in (\ref{r3}) can be expressed
as
\begin{equation}
P_{\nu_e \nu_e} = 1 - 2c_{13}^{2}c_{12}^2 + 2c_{13}^{4}c_{12}^4 - 4
(c_{13}s_{12}c_{23} - s_{13}s_{23})^{2} (c_{13}s_{12}s_{23} +
s_{13}c_{23})^{2}{\sin^2}({\pi L/\lambda_{23}})
\label{pnueatm}
\end{equation}
\begin{equation}
P_{\nu_\mu \nu_e} = 2c_{13}^2c_{12}^2s_{12}^2 - 4c_{12}^2c_{23}s_{23}
(c_{13}s_{12}c_{23} - s_{13}s_{23})(c_{13}s_{12}s_{23} + s_{13}c_{23})
{\sin^2}({\pi L/\lambda_{23}})
\label{pmueatm}
\end{equation}
\begin{equation}
P_{\nu_\mu \nu_\mu} = 1 - 2c_{12}^{2}s_{12}^2 -
4c_{12}^4c_{23}^2s_{23}^2{\sin^2}({\pi L/\lambda_{23}})
\label{pnumuatm}
\end{equation}

The results of the combined analysis of the above three experiments are
presented in figs. 1 and 2 in the large $s_{13}^2$ and small
$s_{13}^2$ limits respectively. It is sufficient to consider these
limits as the allowed values of $s_{13}^2$
are confined in these ranges.
As seen in (\ref{pnumul}), the parametrisation chosen for
the mixing matrix $U$ ensures that the probability relevant for
LSND is independent of the mixing angle $s_{23}$. From the value of
$P_{{\overline{\nu}_{\mu}\overline{\nu}_e}}$ quoted by the LSND group
\cite{lsnd} one can find the allowed area in the $s_{12}^2$ -
$s_{13}^2$ parameter space for fixed values of the ratio
${\Delta{m}^2}L/E$.  The following constraints are found: for $s_{12}$
very small ($\sim$ 0) or very large ($\sim$ 1), $s_{13}$ ranges from
0 $ \leq s_{13} < 1 $ while  for intermediate values of $s_{12}$, only
very large $s_{13}$ values are allowed.  This is between the solid
lines in fig. 1 (2) for large (small) values of $s_{13}$,
in the limit of ${\sin^2}({\pi L/\lambda_{12}}) \sim$
1.  From eq. (\ref{bugey}) the probability for Bugey is also a
function of the same mixing angles $s_{12}$ and $s_{13}$ only, so
that, using their result one can further rule out a portion
of the parameter space -- namely, intermediate $s_{13}$ values at
small $s_{12}$ -- which were allowed by LSND.  This is shown by the
dashed lines in figs. 1 and 2 implying the following
limits for small $s_{12}^2 (<$ 0.0018): either $s_{13}^2 > \sim 0.97$
(fig. 1) or $s_{13}^2 <$ 0.026 (fig. 2). In the other regions of the
parameter space the LSND data puts more severe constraints than Bugey.

Our approach next is to determine how much of the combined allowed
area from LSND and Bugey is consistent with the atmospheric data
for fixed values of $s_{23}$.  The sub-GeV Kamiokande data implies
\begin{equation}
0.48 \leq R \leq 0.73 ~(90\% {\rm C.L.})
\label{r90}
\end{equation}
Imposing this constraint, one finds that the allowed parameter
space (shown
shaded in figs. 1 and 2) depends on the chosen $s_{23}^2$.
In general there are three regions:\\
(i) The large $s_{13}^2 ( > \sim 0.97)$ - small $s_{12}^2 (
< \sim 0.1)$ region shown in fig. 1. In this limit it is the
$\nu_{\mu} - \nu_{e}$ oscillation that dominates. Considering the
limiting case of $s_{12} \rightarrow$ 0 and $s_{13} \rightarrow$ 1,
the relevant probabilities assume the forms:\\
$P_{\nu_e
\nu_e} \simeq 1 - 2 c_{23}^2 s_{23}^2$, $P_{\nu_\mu
\nu_e} \simeq  2 c_{23}^2 s_{23}^2$,  $P_{\nu_\mu \nu_\mu} \simeq
1 - 2 c_{23}^2 s_{23}^2$\\
{}From these expressions it is clear
that in this limit $P_{\nu_\mu \nu_\tau} \simeq$ 0. \\
(ii) The large $s_{13}^2$ and intermediate $s_{12}^2$ zone also
shown in fig. 1. To
understand the transitions in this range we examine the various
probabilities in the limit $s_{13}^2 \rightarrow 1$. In this
limit eqs. (\ref{pnueatm}) - (\ref{pnumuatm}) become \\
$P_{\nu_e \nu_e} \simeq 1 - 2 c_{23}^2 s_{23}^2$,
$P_{\nu_\mu \nu_e} \simeq  2 c_{12}^2 c_{23}^2 s_{23}^2$,
$P_{\nu_\mu \nu_\mu} \simeq 1 - 2c_{12}^{2}s_{12}^2 - 2 c_{12}^4
c_{23}^2s_{23}^2$ \\
This is the region where the depletion can be due to both the
channels simultaneously excepting in the special case of $s_{23}
\rightarrow $ 0 when this
reduces to solely $\nu_{\mu} - \nu_{\tau}$ oscillation.
{}From fig. 1 one also notices
that irrespective of the choice of $s_{23}$, large values of
$s_{12}^2$ around $\sim$ (0.85-1) are disfavoured by the
atmospheric data. In this region $\nu_e - \nu_{\tau}$ conversion is
effective.\\
(iii) The small $s_{12}^2$ -
$s_{13}^2$ zone -- 0 $ < s_{12}^2 < 1.8 \times
10^{-3}$, 0 $\leq s_{13}^2 \leq 0.01$ -- a look at the various
survival and transition probabilities reveals that this is a
region where the depletion is mainly due to $\nu_{\mu} -
\nu_{\tau}$ oscillation.  This can be easily seen by
considering the limiting cases $s_{12}$, $s_{13}$ $\rightarrow$ 0,
when eqs.  (\ref{pnueatm}) - (\ref{pnumuatm}) give
$P_{\nu_e \nu_e} \simeq $ 1, $P_{\nu_\mu
\nu_e} \simeq $ 0, $P_{\nu_\mu
\nu_\mu} \simeq 1 - 2 c_{23}^2 s_{23}^2$. Substituting these in
(\ref{r90}) one finds 0.162 $< s_{23}^2 < $ 0.838. There is a sharp
cut-off as $s_{23}^2$ crosses 0.162 and for practically all
intermediate values upto 0.838, the whole of the parameter space
allowed by LSND and Bugey is consistent with the atmospheric
neutrino data. Thus in this regime we show the allowed region for
only one representative $s_{23}^2$. We have numerically checked
that the allowed region is the same as the one presented in fig. 2
for all other $s_{23}^2$ in the above range.

In our analysis we have fixed $\Delta_{12} \simeq \Delta_{13}$ at $6
eV^{2}$, where LSND is most sensitive and
${\sin^2}({\pi L/\lambda_{12}}) \rightarrow 1$,
maximising the oscillatory term.
As discussed in \cite{silk} it remains
to be seen what best-fit value, consistent with KARMEN and
BNL-E776, emerges when more data is accumulated.
Our results remain unchanged as long as it is
permissible to use the above limit.

For the atmospheric neutrino case we approximate the
${\sin^2}({\pi L/\lambda_{23}})$ factor by its averaged value 0.5
as is often done in the context of the sub-GeV
data \cite{barger,pakvasa,acker}.
This approximation can be improved by an averaging over the
incident neutrino energy  spectrum, the zenith-angle of the beam as
well as the final lepton energy \cite{barger,fogli1}. $r_{MC}$ is
taken to be 0.45 from a detailed Monte-Carlo simulation including
the effects of muon polarisation \cite{gaisser}.

Finally let us discuss the implications of the parameter
space allowed by the Kamiokande atmospheric neutrino, LSND
and Bugey data for the $\nu_{\mu} - \nu_{\tau}$ oscillation search
at CHORUS \cite{chorus} and NOMAD \cite{nomad}. These experiments
use the $\nu_{\mu}$ beam from the CERN SPS with the mean energy
$\sim$ 30 GeV and the approximate
source-detector distance is 0.8 km so that $\lambda_{23} >> L$
and
\begin{equation}
P_{\nu_{\mu} \nu_{\tau}} = 4 c_{12}^2 s_{12}^2 s_{13}^2 \sin^2({\pi
L/\lambda_{13}})
\label{pchorus}
\end{equation}
With the CERN SPS designed to deliver 2.4 $\times 10^{19}$ protons,
CHORUS and NOMAD are sensitive to a minimum oscillation probability of
$10^{-4}$.
For $\Delta_{12} = \Delta_{13}$ in the LSND range of $\sim 6 eV^2$,
$\sin^2({\pi L/\lambda_{13}}) \sim 0.04$, whence (\ref{pchorus}) is
$P_{\nu_{\mu} \nu_{\tau}} \simeq 0.16 c_{12}^2 s_{12}^2 s_{13}^2$.
Then for the three allowed regions in the $s_{12}^2 - s_{13}^2$
plane one gets:\\
(i) In the large $s_{13}^2$, small $s_{12}^2$ zone $P_{\nu_{\mu}
\nu_{\tau}}$ can be slightly greater than $10^{-4}$ being marginally
within the reach of these experiments. This is the $\nu_{\mu} -
\nu_e $ oscillation zone for atmospheric neutrinos.\\
(ii) For large $s_{13}^2$ and intermediate values of $s_{12}^2$,
$P_{\nu_{\mu} \nu_{\tau}}$ is $\sim$ $10^{-2}$, which is well within
the reach of CHORUS and NOMAD. Recall that this is
the genuine three generation oscillation regime for atmospheric
neutrinos where both $\nu_{\mu} - \nu_{\tau}$ and $\nu_{\mu} -
\nu_e$ modes are operative, excepting for the special case of
$s_{23} \simeq$  0 for which it is just $\nu_{\mu} - \nu_{\tau}$.\\
(iii) In the limit of both $s_{12}^2, s_{13}^2$ small,
$P_{\nu_{\mu} \nu_{\tau}}$ is very small and this regime, where the
atmospheric anomaly is due to $\nu_{\mu} - \nu_{\tau}$ oscillation,
cannot be probed by CHORUS and NOMAD.

For the chosen values of the mass-differences a simultaneous
solution to the solar neutrino problem is unobtainable unless one
invokes a sterile neutrino. Work is in progress in this direction
\cite{sg}.

In conclusion, we have obtained the mixing angles
compatible with atmospheric, LSND and reactor experiments (in
particular Bugey) in the context of three flavour mixing.
Keeping ${\Delta{m}}^2$ fixed at the best fit values obtained
from two generation analyses of the LSND and atmospheric
data, we find three regions of parameter space
that can account for all three experiments simultaneously.
Our results differ from an analysis presented in \cite{minakata}
in that we find the
mixing angles to be less restricted.
Our method, which takes into account the possibility that the
depletion of the atmospheric neutrinos can be simultaneously due to
$\nu_{\mu} - \nu_e$ and $\nu_{\mu} - \nu_{\tau}$ oscillations, is
more general and includes the constraints obtained in
\cite{minakata} as a special case. A direct comparison of the
values obtained for the mixing angles is, however, not
proper because the definitions of the mixing matrices are different.
The sensitivity of the accelerator based neutrino oscillation
experiments at CERN,
CHORUS and NOMAD, is different in the three allowed zones
and thus they can
distinguish between these sectors of the
parameter space.
We find that CHORUS and
NOMAD are most sensitive to that part of the combined allowed area
where the
atmospheric neutrino anomaly is due to $\nu_{\mu} - \nu_{\tau}$
and $\nu_{\mu} - \nu_e$ oscillation modes simultaneously.

\vskip 30pt
S.G. is supported by the Council of Scientific and Industrial
Research, India while
A.R. has been supported in part by the Department of Science and
Technology and Council of Scientific and Industrial Research, India.
We thank S. Mohanty for sending us some of the useful literature.
\newpage
\begin{center} {\Large{\bf FIGURE CAPTIONS}}
\end{center}
\vskip 30pt
Figure 1: The 90 \% C.L. allowed region in the $s_{12}^2 -
s_{13}^2$ plane from LSND is between the solid lines, that from
Bugey is above the dashed line while the combined allowed area
including the Kamiokande sub-GeV data is shown shaded. \\  \\
Figure 2: Same as in fig. 1 excepting the region allowed from Bugey is
below the dashed line.\\


\newpage
\begin{thebibliography}{99}
\bibitem{lsnd}W.C. Louis, Nucl. Phys. (Proc. Suppl.) {\bf B38}, 229
(1995); C. Athanassopoulos {\it et al.}, nucl-ex/9504002 (1995).
\bibitem{caldwell}D.O. Caldwell, Nucl. Phys. (Proc. Suppl.) {\bf
B38}, 375 (1995).
\bibitem{primack}J.R. Primack {\it et al.}, Phys. Rev. Lett. {\bf
74}, 2160 (1995); G.M. Fuller, J.R. Primack and Y.Z. Qian,
astro-ph/9502081;
D.O. Caldwell and R.N. Mohapatra, Preprint No. UCSB-HEP-95-1,
hep-ph/9503316.
\bibitem{silk}G. Raffelt and J. Silk, hep-th/9502306.
\bibitem{msw}L. Wolfenstein Phys. Rev. {\bf D34} 969 (1986);
S. P. Mikheyev and A. Yu. Smirnov,  Sov. J. Nucl
Phys. {\bf{42}(6)} 913 (1985); Nuovo Cimento
{\bf 9c}  17 (1986).
\bibitem{langacker}N. Hata and P. Langacker, Phys. Rev.  {\bf D50},
632 (1994).
\bibitem{hata}N. Hata, Univ. of Pennsylvania
Preprint No. UPR-0605T, 1994.
\bibitem{fukuda}Y. Fukuda {\it et al.}, Phys. Lett. {\bf B335}, 237
(1994).
\bibitem{minakata}H. Minakata, Preprint No. TMUP-HEL-9502, March
15, 1995.
\bibitem{fogli1}G.L. Fogli, E. Lisi and D. Montanio, Phys. Rev. {\bf
D49}, 3626 (1994).
\bibitem{imb}D. Casper {\it et al.}, Phys. Rev. Lett. {\bf 66},
2561 (1991); R. Becker-Szendy {\it et al.}, Phys. Rev. {\bf D46},
3720 (1992).
\bibitem{hirata}K. S. Hirata {\it et al.}, Phys. Lett. {\bf B280}, 146
(1992).
\bibitem{fogli2} The usage of the `ratio of ratios' R as a valid
indicator of the neutrino anomaly has been recently critically
examined. see G.L. Fogli and E. Lisi, Institute for Advanced Study
report IASSNS-AST 95/21 (unpublished).
\bibitem{cern}D. Saltzberg, Report no. hep-ph-9504343 (unpublished).
\bibitem{karmen}B. Seligmann in Proc. of the XXVII Int. Conf. on High
Energy Physics (Glasgow), Eds. P.J. Bussey and I.G Knowles, Inst. of
Phys. Publising, Bristol, (1995) p683; G. Drexlin, Prog. in Part.
and Nucl. Phys., {\bf 32}, 375 (1994);
B. Armbruster {\it et al.} (KARMEN Collaboration), Nucl. Phys.
(Proc. Suppl.) {\bf B38}, 235 (1995).
\bibitem{bnl}L.\ Borodovsky {\it et al.}, Phys.\ Rev.\ Lett. {\bf
68}, 274 (1992).
\bibitem{bugey}B. Achkar {\it et al.}, Nucl. Phys.  {\bf B434},
503 (1995).
\bibitem{barger}V. Barger and K. Whisnant, Phys. Lett. {\bf B209},
365 (1988).
\bibitem{pakvasa}A. Acker, J.G. Learned, S. Pakvasa and T.J.
Weiler, Phys. Lett. {\bf B298}, 149 (1993).
\bibitem{acker}A. Acker, A.B. Balantekin and F. Loreti, Preprint No.
Mad/NT/93-07, MAD/PH/\# 774, July (1993).
\bibitem{gaisser}G. Barr, T.K. Gaisser and T. Stanev, Phys. Rev
{\bf D39} 3532 (1989); T.K. Gaisser, T. Stanev and G. Barr, {\it
ibid} {\bf D38}, 85 (1988).
\bibitem{chorus}M. de Jong {\it et al.}, CERN-PPE/93-131 (1993);
N. Armenise {\it et al.}, CERN-SPSC/90-42 (1990).
\bibitem{nomad}P. Astier {\it et al.}, CERN-SPSLC/91-21,
CERN-SPSLC/91-48, CERN-SPSLC/P261 Add.1 (1991).
\bibitem{sg}S. Goswami, in preparation.
\end{thebibliography}
\end{document}